\documentclass{article}
\usepackage{spconf,amsmath,graphicx}
\usepackage{stfloats}
\usepackage{ragged2e}
\usepackage{amsmath}

\renewcommand{\justify}{\leftskip=0pt \rightskip=0pt plus 0cm}

\title{Memory-augmented Contrastive Learning for Talking Head Generation}
\name{Jianrong Wang$^{1}$ \qquad Yaxin Zhao$^{2}$ \qquad Li Liu$^{3\star}$\thanks{$^{\star}$Corresponding author: avrillliu@hkust-gz.edu.cn} \qquad Hongkai Fan$^{1}$ \qquad Tianyi Xu$^{1}$ \qquad Qi Li$^{4}$ \qquad Sen Li$^{1}$ }
\address{
  $^{1}$College of Intelligence and Computing, Tianjin University, Tianjin, China\\
  $^{2}$Tianjin International Engineering Institute, Tianjin University, Tianjin, China\\
  $^{3}$The Hong Kong University of Science and Technology (Guangzhou), Guangzhou, China\\
  $^{4}$School of Electrical and Information Engineering, Tianjin University, Tianjin, China
  }

\begin{document}
%
\maketitle
\begin{abstract}
Given one reference facial image and a piece of speech as input, talking head generation aims to synthesize a realistic-looking talking head video. However, generating a lip-synch-\\ronized video with natural head movements is challenging. The same speech clip can generate multiple possible lip and head movements, that is, there is no one-to-one mapping relationship between them. To overcome this problem, we propose a \emph{S}peech \emph{F}eature \emph{E}xtractor (SFE) based on memory-augmented self-supervised contrastive learning, which introduces the memory module to store multiple different speech mapping results. In addition, we introduce the \emph{M}ixed \emph{D}ensity \emph{N}etworks (MDN) into the landmark regression task to generate multiple predicted facial landmarks. Extensive qualitative and quantitative experiments show that the quality of our facial animation is significantly superior to that of the state-of-the-art (SOTA). The code has been released at https://github.com/Yaxinzhao97/MACL.git.
\end{abstract}

\vspace{-0.4cm}
\section{Introduction}
\label{sec:intro}
\vspace{-0.3cm}

\begin{sloppypar}
Talking head generation is crucial to film making \cite{kim2019neural}, audio-visual speech generation \cite{liu2020re} \cite{liu2018visual}, computer games \cite{htike2017review}, and so on. Head pose plays an important role in enhancing human perception of the authenticity of generated video \cite{chen2020talking}. 
\end{sloppypar}
It was reported that literature \cite{DBLP:ChenLMDX18} \cite{song2018talking} \cite{DBLP:VougioukasPP20} directly mapped speech to talking head videos. \cite{DBLP:ChenMDX19} \cite{zhou2020makelttalk} first established the mapping from speech to high-level representations, \emph{i.e.}, the facial landmarks, and then generated talking head videos based on the landmarks. Compared with the face pixel image, the facial landmarks are relatively sparse, so we can pay more attention to capturing the speaker's head movements.

At present, supervised learning is the mainstream method of speech feature extraction in talking head generation. Supervised learning method requires large effort to label the data and the accuracy of the data label cannot be guaranteed, which makes speech feature extraction easy to be affected by noisy data. Therefore, we propose to use the self-supervised \cite{wang2021self} method to extract speech features. Then, the learned speech features are applied to the downstream task for fine-tuning, so as to obtain speech features that is more consistent with the downstream task.

Another challenge is that the mapping of speech to lip and head movements is not one-to-one. For example, when we say ``Hi", the degree of mouth opening may be different with different people. Similarly, the head movements may also vary. We propose to deal with this uncertainty in two stages. In the stage of speech features extraction, the memory module is introduced into the self-supervised contrastive learning \cite{chen2021uscl}. Multiple results generated when speech is mapped to lip and head movements are allocated to the memory module so that the feature extractor can focus on speech feature extraction. For uncertainty problems, using a single model will always lead to sub-optimal predictions \cite{varamesh2020mixture}. Therefore, in the  stage of facial landmark regression, we introduce the mixture density networks to generate multiple predicted landmarks.

The contributions of our work can be summarized as:
\begin{enumerate}
\item We propose a speech feature extraction network based on memory-augmented self-supervised contrastive learning to extract speech features. 
\item We introduce the mixture density networks into facial landmarks regression task to generate multiple predicted facial landmarks, which can improve the naturalness of generating head movements.
\item Experimental results in multi-speaker scenarios show that our model is superior to SOTA in terms of landmarks generation and rhythmic head movements.

\end{enumerate}

\begin{figure*}[hbpt]
\centering
\includegraphics[width=0.9\textwidth]{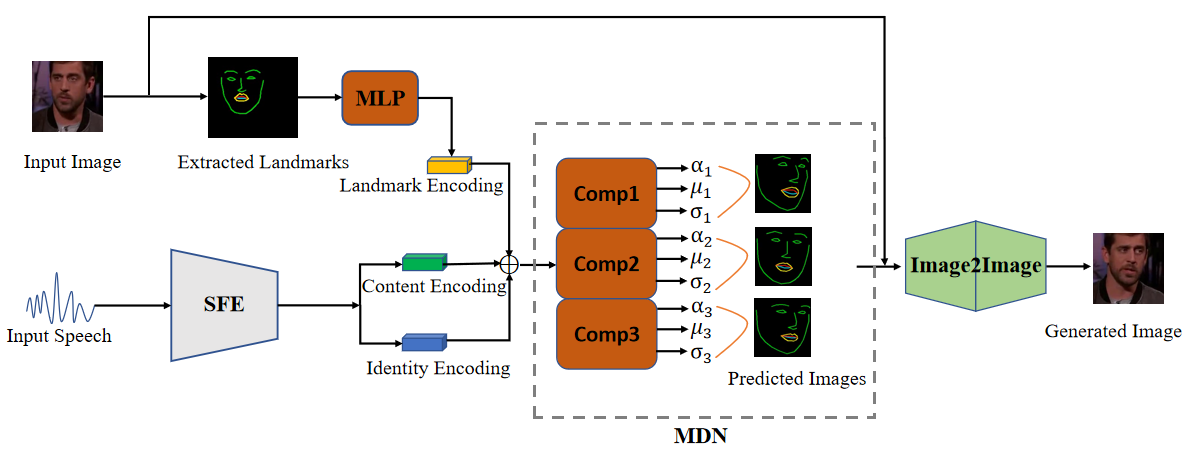}
\caption{Overview of the proposed model. First, we extract the content and identity features from the speech signal through the SFE. Then we input the extracted features and reference facial landmark features into MDN network to predict facial landmarks. Each component is a simple multi-layer perceptron. Finally, facial landmarks are converted into photo-realistic facial images through \emph{Image2Image} network.}
\label{fig:overall}
\end{figure*}

\vspace{-0.5cm}
\section{Method}
\vspace{-0.2cm}

\justify
As shown in Fig. \ref{fig:overall}, our model consists of three key components. The first is \emph{Speech Feature Extractor} (SFE), which extracts content and identity information from speech signals. The second is \emph{Mixed Density Networks} (MDN), which takes the speech features and the landmark features of the reference face as input, predicts a variety of possible facial landmarks, and finally selects the one with the best prediction result as the output. The last part is the \emph{Image2Image} network, which translates facial landmarks into photo-real facial images. We describe each module of our algorithm in the following subsections.

\vspace{-0.2cm}
\begin{sloppypar}
\subsection{Speech Feature Extractor Based on Memory-augmented Contrastive Learning}
\end{sloppypar}

\begin{figure}[htbp]
\centering
\setlength{\abovecaptionskip}{-0.5cm}
\includegraphics[width=0.5\textwidth]{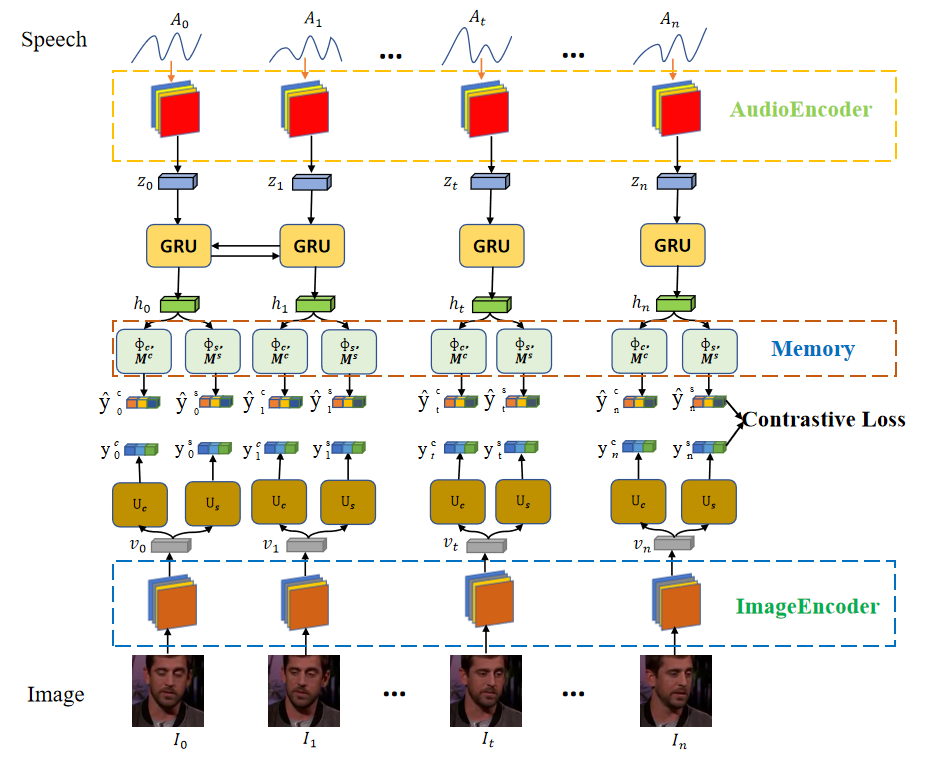}
\caption{Overall architecture of the pre-trained SFE. Content features and identity features are extracted from speech and image respectively, and the corresponding contrast loss is calculated. Finally, take the upper part, the audio module, as the audio feature extractor.}
\label{fig:selfStruc}
\end{figure}

The proposed SFE mainly contains three parts: audio encoder, image encoder and memory module, as shown in Fig.~\ref{fig:selfStruc}.

\textbf{Memory Module.} To deal with the one to many mapping problem of speech to lip and head movements, we introduce memory module \cite{han2020memory} into self-supervised contrastive learning to store the multiple mapping results. 

In detail,  the content memory module is denoted as $M^{c} ={[m_{1}^c,m_{2}^c, ... ,m_{k}^c]}^{T} \in R^{k \times C}$, where $k$ is the size of memory module and $C$ is the dimension of each memory slot. The weight of each memory slot was obtained by the content probability distribution function $\phi_{c}$, and then the predicted speech content feature $\hat{y}_{t}^c$ was calculated. The formulas are as follows:
\begin{equation}\label{equ:paddress}
\setlength{\abovedisplayskip}{5pt} 
\setlength{\belowdisplayskip}{5pt} 
\begin{split}
p_{t}^c = Softmax(\phi_{c}(h_{t})),
\end{split}
\end{equation}
\begin{equation}\label{equ:memcontent}
\begin{split}
\hat{y}_{t}^c = \sum_{i=1}^{k} p_{i, t}^c \cdot m_{i}^c = p_{t}^{c}M^c,
\end{split}
\end{equation}

where $p_{i, t}^c \in R^{k}$ is the contribution of the $i$-th content memory slot for the content feature representation at time step $t$. $h_{t}$ is the speech feature extracted by Bi-GRU at time step $t$.  

The calculation method of speech identity features $\hat{y}_{t}^s$ and content features are similar, but $h_{t}$ obtains the identity weight vector through the identity weight distribution function $\phi_{s}$, and then performs the product operation with the identity memory module $M^s$.

\textbf{Contrastive Loss Functions.} By calculating the distance between the speech features and the image features, the distance lip-synced speech-image pair is smaller than unsynchronized speech-image pair. Identity contrast loss function and content contrast loss function are similar in form. The contrastive loss function between a group of positive sample pairs are as follows:

\begin{equation}\label{equ:lc}
\setlength{\abovedisplayskip}{-5pt}
\begin{split}
L_c = -\frac{1}{T}\sum_{t=1}^{T} log\frac{S(\hat{y}_{t}^c, y_t^c)}{\sum_{k=1}^{T} S(\hat{y}_{t}^c, y_k^c)},
\end{split}
\end{equation}
\begin{equation}\label{equ:memcontent}
\begin{split}
S(\hat{y}_{t}^c, y_t^c) = exp(w \cdot cos(\hat{y}_{t}^c, y_t^c) + b),
\end{split}
\end{equation}
where $\hat{y}_{t}^c$ is the content feature representation of speech, $y_t^c$ is the content feature representation of image sequence, $t$ is the number of frames, and $w$ and $b$ are learnable parameters.

\vspace{-0.2cm}
\subsection{Facial Landmarks Regression based on Mixture Density Networks}

Most existing works generated single facial landmark by minimizing the negative log likelihood of a single Gaussian distribution, i.e., the mean square error. However, mapping from speech to head movements and lip movements has significant uncertainty, so previous works has bottlenecked the accuracy improvement of face landmarks prediction. For uncertain prediction tasks, mixture density networks is a powerful tool. This work proposes to estimate multiple facial landmarks by minimizing the negative log likelihood of mixed Gaussian distribution. The output of the $MDN$ model is a set of mixing coefficients $\alpha$ and Gaussian kernel parameters (i.e., mean $\mu$ and variance $\sigma$). The mean of Gaussian kernel is the generated facial landmark, and the mixing coefficient and variance represent the uncertainty of each generated predicted facial landmark.

Different from the previous methods, instead of directly generating the final target facial landmarks, we first generate the aligned target facial landmarks $p_{align}$, rotation matrix $R_r$ and offset matrix $R_t$, and then use the inverse affine transformation to get the final target facial landmarks, i.e., by subtracting $R_t$ from $p_{align}$ and then multiplying the results with $R_r$. This reduces the predictive pressure on the identity features in the facial landmarks.


\textbf{Model Inference.} We can use either the mixture of outputs by the components or the one with the highest score to generate the facial landmarks parameters. We try both cases and found out that using the maximum component leads to slightly better results.

\vspace{-0.3cm}
\subsection{Image2Image} 
Finally, we input target facial landmarks and reference facial image to generate the final target facial image. We use the image-to-image translation module in \cite{zhou2020makelttalk} and fine-tune it.

\section{Experiment and analysis}
\label{sec:typestyle}

\vspace{-0.1cm}
\subsection{Implementation Details}
We train both the speech feature extraction model and facial landmarks regression model on VoxCeleb2 \cite{DBLP:ChungNZ18}. It contains more than 1M utterances from over 6,000 celebrities collected from around 150K videos on Youtube. The dataset is fairly gender-balanced (61$\%$ male). 
 Our network is implemented using PyTorch. During the training, we used Adam optimizer, and all models are trained and tested on a single NVIDIA Tesla V100. For the image stream, we first extract all the images in the video, and then extract the landmarks. As for speech data, the sampling rate is 16KHz. We extract fbank at the window size of 20ms and hop size of 10ms.

\vspace{-0.3cm}
\subsection{Comparison with SOTA} 

We extract identity embeddings for the data in the VoxCeleb1 \cite{nagrani2017voxceleb} test set, and evaluate using the self-supervised embeddings directly (without any fine-tuning). We compare the results with those published in the baseline model \cite{nagrani2020disentangled}. As shown in Table \ref{tab:speakerVeri}, our results are 1.05\% lower than the best result in [18] in terms of the EER evaluation metric, which indicates that our approach extracts the speech recognition feature vector effectively.

\vspace{-0.4cm}
\begin{table}[htbp]
\centering
\small\addtolength{\tabcolsep}{20pt}
\caption{Speaker verification results under VoxCeleb1 test set. $\downarrow$ means the lower the better.}
\label{tab:speakerVeri}
\begin{tabular}{c | c}
\hline
 \textbf{Method} & \textbf{EER(\%)$\downarrow$}  \\
\hline
 IL only \cite{nagrani2020disentangled} & 23.15 \\
\hline
 IL + CL \cite{nagrani2020disentangled}& 22.59 \\
\hline
IL + Disent. loss \cite{nagrani2020disentangled} &  22.09 \\
\hline
Ours & \textbf{21.04} \\
\hline
\end{tabular}
\end{table}

We compare our model with two SOTA methods \cite{DBLP:ChenMDX19} \cite{zhou2020makelttalk}. The quantitative results are illustrated in Table \ref{tab:eval}. We use LMD \cite{chen2018lip} (Landmark Distance) to measure whether lip sounds are synchronized, and RD (Rotation Distance) to measure whether the generated face video has natural head movements. To further evaluate the quality of the generated images of different methods, we compare the SSIM \cite{wang2004image} and PSNR \cite{narvekar2009no}. Although MakitTalk exceeds the proposed method in SSIM and PSNR, it is lower than the proposed method in generation speed (fps). Note that the image resolution generated by MakeitTalk and our method is 256 × 256, while that generated by ATVGNet is 128 × 128.

\vspace{-0.4cm}
\begin{table}[htbp]
\caption{Evaluation results under VoxCeleb2 test set. $\uparrow$ means the higher the better.}
\label{tab:eval}
\scalebox{0.9}{
    \begin{tabular}{c | c | c | c | c | c}
    \hline
     \textbf{Method} & \textbf{LMD$\downarrow$} &  \textbf{RD$\downarrow$} & \textbf{SSIM$\uparrow$} &  \textbf{PSNR$\uparrow$} & \textbf{FPS$\uparrow$}\\
    \hline
    ATVGNet \cite{DBLP:ChenMDX19} & 2.12 &  0.21 & 0.81 &  28.14 & \textbf{34.53}\\
    \hline
    MakeitTalk \cite{zhou2020makelttalk} & 7.85 & 0.08  & \textbf{0.83} & \textbf{29.77} & 22\\
    \hline
    Ours & \textbf{1.83} & \textbf{0.07}  & \textbf{0.83} & 28.91 & 25.30\\
    \hline
    \end{tabular}
}
\end{table}

\begin{figure}[htbp]
\centering
\setlength{\abovecaptionskip}{-0.5cm}
\includegraphics[width=0.40\textwidth]{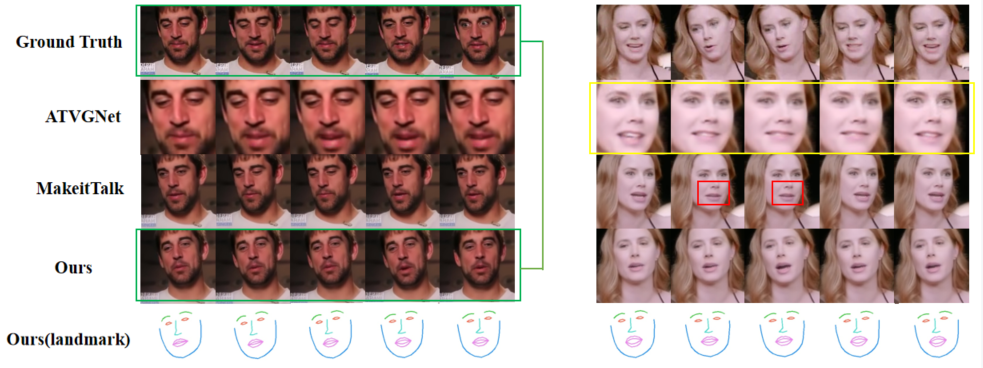}
\caption{\textbf{Qualitative results produced by ATVGNet \cite{DBLP:ChenMDX19}, MakeitTalk \cite{zhou2020makelttalk} and Ours.} ATVGNet generated facial animation without head movements (as shown in the yellow box), The lip shape generated by MakeitTalk is inaccurate (as shown in the red box). Our results have a more pronounced trend in head movement (as shown in the green box).}
\label{fig:ourres}
\end{figure}

As shown in Fig. \ref{fig:ourres}, ATVGNet only focused on generating facial images without generating natural head movements. While MakeitTalk generates subtle head movements, the mouth shapes of their model are not accurate. Compared with these methods, our method generates more natural and lip-synchronized facial animation.

\begin{figure}[htbp]
\centering
\includegraphics[width=0.45\textwidth]{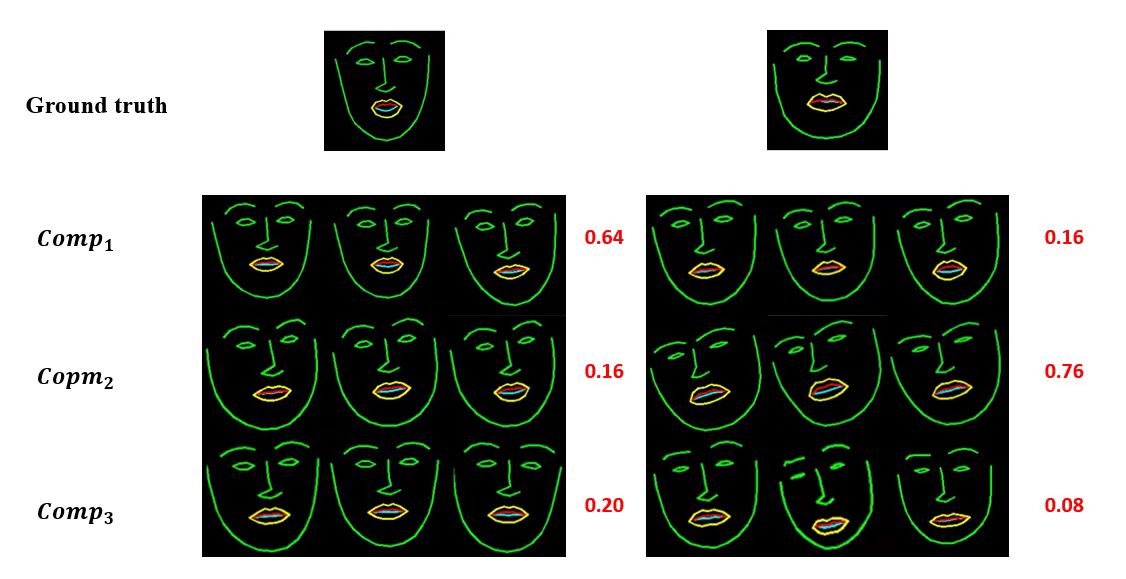}
\caption{\textbf{MDNs visualize results for each component.} The red numbers on the right are the mixing coefficient corresponding to each mixing component.}
\label{fig:mdnvis}
\end{figure}

We visualize the generated results of each component to study the features learned by each mixture component, as shown in Fig. \ref{fig:mdnvis}. The main difference between each component is the learned features related to the speaker's identity, including facial contour shapes and head movements. $Comp_{1}$ learned thin and narrow facial contours, $Comp_{2}$ learned wide and fat facial contours, and $Comp_{3}$ learned random facial contours. Among them, the facial contour generated by the component with the largest mixing probability $\alpha_{m}$ is closest to the real speaker.

\vspace{-0.4cm}
\subsection{Ablation Study} 
For self-supervised contrastive learning, we conduct ablation experiments to study the contributions of the memory module in our full model. For facial landmarks regression model, we conduct ablation experiments to study the contributions of the two components in our full model separately: number of mixture components and facial landmarks regression method. 

\textbf{Memory Module.} As shown in Table \ref{tab:memAblation}, by comparing the experimental results with no memory module Ours (wo), only a single memory module Ours (w), using identity memory module and content memory module Ours (cs), it is shown that the memory module has a great effect on the quality of feature extraction. 

\begin{table}[htbp]
\centering
\small\addtolength{\tabcolsep}{9pt}
\caption{Comparison of results of memory module ablation experiment in speaker verification.}
\label{tab:memAblation}
\begin{tabular}{c | c | c | c }
\hline
 \textbf{Method} & \textbf{EER(\%)$\downarrow$} & \textbf{LMD$\downarrow$} & \textbf{RD$\downarrow$}\\
\hline
 Ours (wo)& 22.12 & 0.14 & 0.08 \\
\hline
 Ours (w)& 21.46  & 0.13 & 0.07\\
\hline
Ours (cs) &  \textbf{21.04} & \textbf{0.12} & \textbf{0.07}\\
\hline
\end{tabular}
\end{table}

\textbf{Number of Mixture Components.} It can be seen from Table  \ref{tab:numMLMD}. For LMD, when the number of mixture components $M$ is greater than 3, its value basically does not decrease with the increase of $M$. As for RD, when the number of mixture components $M$ is greater than 5, it will not decrease basically. This indicates that each component mainly learns information related to the speaker's identity, and more components can model more different speakers.

\vspace{-0.4cm}
\begin{table}[htbp]
\centering
\small\addtolength{\tabcolsep}{1pt}
\caption{The influence of the number of mixed components on LMD.}
\label{tab:numMLMD}
\begin{tabular}{c | c | c | c | c | c}
\hline
\textbf{M} & \textbf{1} & \textbf{2} & \textbf{3} & \textbf{5} & \textbf{8} \\
\hline
LMD$\downarrow$ & 0.14 & 0.13 & \textbf{0.12} & \textbf{0.12} & \textbf{0.12} \\
\hline
RD$\downarrow$ & 0.08 & 0.08 & 0.07 & \textbf{0.06} & \textbf{0.06} \\
\hline
\end{tabular}
\end{table}

\begin{figure}[htbp]
\centering
\includegraphics[width=0.47\textwidth]{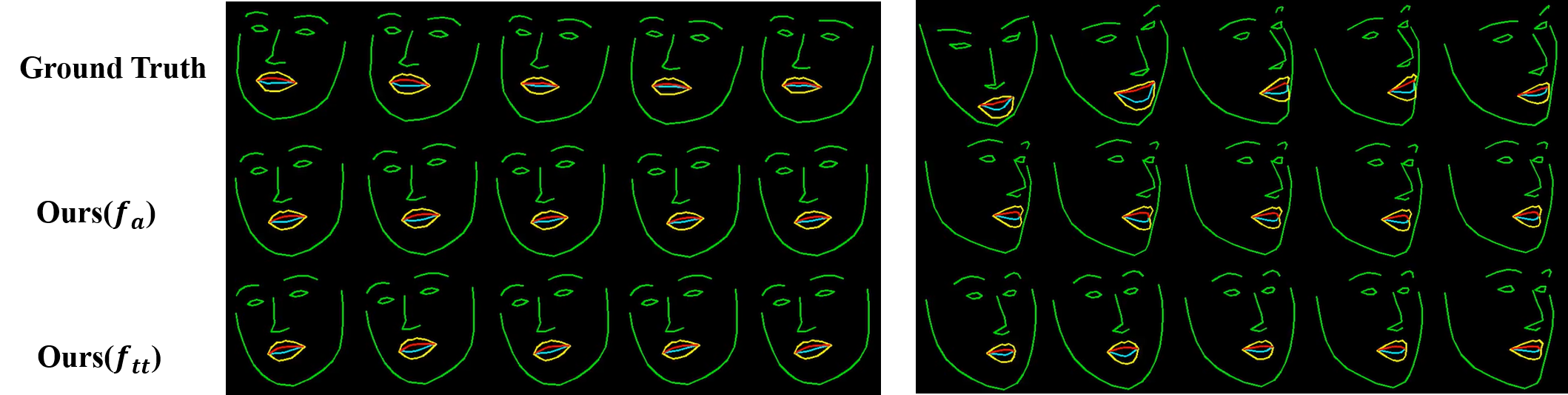}
\setlength{\abovecaptionskip}{-0.1cm}
\caption{Illustration of different facial landmark regression method results. $f_a$ represents direct regression to facial landmarks, $f_{tt}$ represents that we first generate facial landmark, rotation matrix and offset matrix aligned with the standard face, and then obtain the facial landmark through the inverse affine transformation.}
\label{fig:diffMethodAblation}
\end{figure}

\textbf{Facial Landmarks Regression Method.} In Fig. \ref{fig:diffMethodAblation}. Ours ($f_{tt}$) can generate more natural head movements for the source video with large head movements on the right, while Ours ($f_a$) can only generate slight head movements. Although our method does not generate exactly the same head movements as ground truth, the dynamic trends are similar.

\vspace{-0.8cm}
\section{Conclusions}
\label{sec:majhead}
\vspace{-0.2cm}

By leveraging self-supervised constrastive learning, memory module and mixture density networks, the proposed method can synthesize talking head videos with natural head movements. Future work will focus on extracting emotion embeddings from speech and adding emotion embeddings to talking head generation.

\vspace{-0.3cm}
\section{Acknowledgement}

This work is supported by the National Natural Science Foundation of China (No. 61977049), the National Natural Science Foundation of China (No. 62101351), and the GuangDong Basic and Applied Basic Research Foundation (No.2020A1515110376).


\begin{thebibliography}{10}

\bibitem{kim2019neural}
Hyeongwoo Kim, Mohamed Elgharib, Michael Zollh{\"o}fer, Hans-Peter Seidel,
  Thabo Beeler, Christian Richardt, and Christian Theobalt,
\newblock ``Neural style-preserving visual dubbing,''
\newblock {\em ACM Transactions on Graphics (TOG)}, vol. 38, pp. 1--13, 2019.

\bibitem{liu2020re}
Li~Liu, Gang Feng, Denis Beautemps, and Xiao-Ping Zhang,
\newblock ``Re-synchronization using the hand preceding model for multi-modal
  fusion in automatic continuous cued speech recognition,''
\newblock {\em IEEE Transactions on Multimedia}, vol. 23, pp. 292--305, 2020.

\bibitem{liu2018visual}
Li~Liu, Thomas Hueber, Gang Feng, and Denis Beautemps,
\newblock ``Visual recognition of continuous cued speech using a tandem cnn-hmm
  approach,''
\newblock in {\em Interspeech}, 2018, pp. 2643--2647.

\bibitem{htike2017review}
Kyaw~Kyaw Htike,
\newblock ``A review on data-driven learning of a talking head model,''
\newblock {\em International Journal of Intelligent Systems Technologies and
  Applications}, vol. 16, pp. 169--190, 2017.

\bibitem{chen2020talking}
Lele Chen, Guofeng Cui, Celong Liu, Zhong Li, Ziyi Kou, Yi~Xu, and Chenliang
  Xu,
\newblock ``Talking-head generation with rhythmic head motion,''
\newblock in {\em Proceedings of the European Conference on Computer Vision
  (ECCV)}, 2020, pp. 35--51.

\bibitem{DBLP:ChenLMDX18}
Lele Chen, Zhiheng Li, Ross~K. Maddox, Zhiyao Duan, and Chenliang Xu,
\newblock ``Lip movements generation at a glance,''
\newblock in {\em Proceedings of the European Conference on Computer Vision
  (ECCV)}, 2018, pp. 538--553.

\bibitem{song2018talking}
Yang Song, Jingwen Zhu, Dawei Li, Xiaolong Wang, and Hairong Qi,
\newblock ``Talking face generation by conditional recurrent adversarial
  network,''
\newblock {\em arXiv preprint: 1804.04786}, 2018.

\bibitem{DBLP:VougioukasPP20}
Konstantinos Vougioukas, Stavros Petridis, and Maja Pantic,
\newblock ``Realistic speech-driven facial animation with gans,''
\newblock {\em International Journal of Computer Vision}, vol. 128, pp.
  1398--1413, 2020.

\bibitem{DBLP:ChenMDX19}
Lele Chen, Ross~K. Maddox, Zhiyao Duan, and Chenliang Xu,
\newblock ``Hierarchical cross-modal talking face generation with dynamic
  pixel-wise loss,''
\newblock in {\em Proceedings of the IEEE/CVF Conference on Computer Vision and
  Pattern Recognition (CVPR)}, 2019, pp. 7832--7841.

\bibitem{zhou2020makelttalk}
Yang Zhou, Xintong Han, Eli Shechtman, Jose Echevarria, Evangelos Kalogerakis,
  and Dingzeyu Li,
\newblock ``Makelttalk: speaker-aware talking-head animation,''
\newblock {\em ACM Transactions on Graphics (TOG)}, vol. 39, pp. 1--15, 2020.

\bibitem{wang2021self}
Jianrong Wang, Ge~Zhang, Zhenyu Wu, Xuewei Li, and Li~Liu,
\newblock ``Self-supervised depth estimation via implicit cues from videos,''
\newblock in {\em IEEE International Conference on Acoustics, Speech and Signal
  Processing (ICASSP)}, 2021, pp. 2485--2489.

\bibitem{chen2021uscl}
Yixiong Chen, Chunhui Zhang, Li~Liu, Cheng Feng, Changfeng Dong, Yongfang Luo,
  and Xiang Wan,
\newblock ``Uscl: pretraining deep ultrasound image diagnosis model through
  video contrastive representation learning,''
\newblock in {\em Medical Image Computing and Computer Assisted Intervention
  (MICCAI)}, 2021, pp. 627--637.

\bibitem{varamesh2020mixture}
Ali Varamesh and Tinne Tuytelaars,
\newblock ``Mixture dense regression for object detection and human pose
  estimation,''
\newblock in {\em Proceedings of the IEEE/CVF Conference on Computer Vision and
  Pattern Recognition (CVPR)}, 2020, pp. 13086--13095.

\bibitem{han2020memory}
Tengda Han, Weidi Xie, and Andrew Zisserman,
\newblock ``Memory-augmented dense predictive coding for video representation
  learning,''
\newblock in {\em Proceedings of the European Conference on Computer Vision
  (ECCV)}, 2020, pp. 312--329.

\bibitem{DBLP:ChungNZ18}
Joon~Son Chung, Arsha Nagrani, and Andrew Zisserman,
\newblock ``Voxceleb2: Deep speaker recognition,''
\newblock in {\em Annual Conference of the International Speech Communication
  Association (ISCA)}, 2018, pp. 1086--1090.

\bibitem{nagrani2017voxceleb}
Arsha Nagrani, Joon~Son Chung, and Andrew Zisserman,
\newblock ``Voxceleb: a large-scale speaker identification dataset,''
\newblock {\em arXiv preprint: 1706.08612}, 2017.

\bibitem{nagrani2020disentangled}
Arsha Nagrani, Joon~Son Chung, Samuel Albanie, and Andrew Zisserman,
\newblock ``Disentangled speech embeddings using cross-modal
  self-supervision,''
\newblock in {\em IEEE International Conference on Acoustics, Speech and Signal
  Processing (ICASSP)}, 2020, pp. 6829--6833.

\bibitem{chen2018lip}
Lele Chen, Zhiheng Li, Ross~K Maddox, Zhiyao Duan, and Chenliang Xu,
\newblock ``Lip movements generation at a glance,''
\newblock in {\em Proceedings of the European Conference on Computer Vision
  (ECCV)}, 2018, pp. 520--535.


\bibitem{wang2004image}
Zhou Wang, Alan~C Bovik, Hamid~R Sheikh, and Eero~P Simoncelli,
\newblock ``Image quality assessment: from error visibility to structural
  similarity,''
\newblock {\em IEEE transactions on image processing}, vol. 13, pp. 600--612,
  2004.

\bibitem{narvekar2009no}
Niranjan~D Narvekar and Lina~J Karam,
\newblock ``A no-reference perceptual image sharpness metric based on a
  cumulative probability of blur detection,''
\newblock in {\em International Workshop on Quality of Multimedia Experience},
  2009, pp. 87--91.

\end{thebibliography}

\end{document}